\documentstyle[epsfig]{mn}
\def\Msun{\ifmmode{~{\rm M}_\odot}\else${\rm M}_\odot$~\fi}
\def\kms{\ifmmode{$~km\thinspace s$^{-1}}\else km\thinspace s$^{-1}$\fi}

\title[Luminosity-metallicity relation for stars on the lower main
sequence]{Luminosity-metallicity relation for stars on the lower main
sequence} \author[Eira Kotoneva, Chris Flynn and Raul Jimenez]
{Eira Kotoneva$^1$, Chris Flynn$^{1,2}$ and Raul Jimenez$^3$ \\ 
$^1$Tuorla Observatory, Piikki\"o, FIN-21500, Finland; eianko@astro.utu.fi; cflynn@astro.utu.fi\\
$^2$Centre for Astrophysics and Supercomputing, Swinburne University of 
Technology, Hawthorn, Australia\\
$^3$Department of Physics and Astronomy, Rutgers University, NJ-08854-8019, USA; raulj@physics.rutgers.edu}
\begin{document}
\maketitle

\voffset=-1.0cm     % standard value

\begin{abstract}

We present a comparison of the predictions of stellar models with the
luminosity of the lower main sequence ($5.5 < M_V < 7.3$) using K dwarfs in the
Hipparcos catalog. The parallaxes of our comparison stars are known to better
than 15\% and metallicities have been recently determined from photometry. A
major advantage of our comparison is that distances in our sample are known
with good accuracy, while tests that involve open and globular clusters are
constrained by potentially inaccurate distances. We show that the luminosity of
the lower main sequence relative to a fiducial (solar metallicity) isochrone is
a simple function of metallicity: $\Delta M_V = 0.84375 \times {\rm [ Fe/H]} -
0.04577$.  We compare the data with a range of isochrones from the literature.
None of the models fit all the data, although some models do clearly better
than others. In particular, metal rich isochrones seem to be difficult to
construct. 

The relationship between luminosity, colour and metallicity for K dwarfs is
found to be very tight.  We are thus able to derive metallicities for K dwarfs
based on their position in the Hipparcos colour-magnitude diagram with
accuracies better than 0.1 dex.  The metallicity-luminosity relation for K
dwarfs leads to a new distance indicator with a wide range of possible
applications.

\end{abstract}


\section{Introduction}

Stellar isochrones are commonly used to predict or interpret the properties of
distant stellar systems. Two popular uses are to derive ages and metallicities
of resolved stellar populations (such as open and globular clusters) and to
determine the evolutionary state and ages of unresolved systems using the
integrated light and synthetic stellar population models (see e.g.  Stetson,
VandenBerg \& Bolte 1996; Sarajedini, Chaboyer \& Demarque 1997; Jimenez,
Flynn \& Kotoneva 1998; Carraro, Girardi \& Chiosi 1999; Chaboyer, Green \&
Liebert 1999; Liu \& Chaboyer 2000; Jimenez 1999). The agreement between
stellar theoretical models and the Sun is outstanding (e.g Bahcall et al
2001), but comparisons of models to data for other metallicities and masses is
still an arduous task.  Comparisons of this type have been restricted in the
past to open and globular clusters (e.g. Westera et al 2002, Cassisi et al
2000), but in both cases distances and luminosities are a source of
considerable uncertainty.  Nevertheless, good agreement has been found between
isochrones and data from globular clusters, i.e. isochrones at low metallicity
are good. Much more difficult to check is the accuracy of metal rich
isochrones ([Fe/H] $ \ga -1$).

Our aim in this paper is to make such a comparison for stars on the lower main
sequence (K dwarfs), using the accurate distances (and thus luminosities)
provided by the ESA Hipparcos mission. The width of the lower main sequence has
long been held to be a consequence of the metallicities of the stars (for a
review see e.g. Reid, 1999). However, this has been difficult to test without
having precise parallaxes and metallicities and furthermore the ability to
remove multiple stars. The Hipparcos catalogue makes this possible for the
first time. In this paper we use Hipparcos parallaxes of a sample of 213 nearby
K dwarfs (in the absolute magnitude range $5.5 < M_V < 7.3$, or broadly G8 to
K3), for which photometric metallicities are available from Kotoneva and Flynn
(2002), to calibrate the luminosity of K dwarfs of a given colour as a function
of metallicity and compare these luminosities to isochrones from the
literature.

The paper is organised as follows. In section 2 we describe the sample. In
section 3 we compare different isochrone sets from the literature to stars of
similar metallicity in the sample. Having found that no isochrone set is able
to fit the whole range of metallicities in the data, we construct an empirical
calibration of the main sequence luminosity with the metallicity in section 4.
Comparing with the original sample of spectroscopically analysed K dwarfs, the
relation turns out to be surprisingly precise, and we discuss applications of
this calibration in section 5. We draw our conclusions in section 6.

\section{The sample}

We have chosen all stars in the survey part of the Hipparcos output catalogue
in the absolute magnitude $M_V$ range $5.5 < M_V < 7.3$.  We term these stars
{\it K dwarfs}. The survey part of the Hipparcos catalogue is complete to
apparent visual magnitude $V < 7.3 + 1.1{\mathrm sin}|b|$, where $b$ is the
galactic latitude (the apparent visual magnitude limit was made dependent on
$b$ in order to avoid observing excessive numbers of stars in the Galactic
plane) (ESA, 1997). This $V$ magnitude limit results in 209 stars. To augment
the basic sample we increased the magnitude limit by 0.9 mag, i.e. the limit is
$V < 7.3 + 1.1{\mathrm sin}|b|$ + $0.9$. This resulted in 668 K dwarfs. For all
stars the parallax, proper motions and $B - V$ colour and apparent $V$
magnitude are available in the Hipparcos catalogue (for more details about the
sample and the observations see Kotoneva and Flynn, 2002).

\subsection{Multiple Stars}

The photometrically determined metallicities for our K dwarfs is only
appropriate for single stars. Identifying the probable multiple stars was
therefore very important. The Hipparcos catalog includes a flag for so called
``probable multiple stars'', based on the quality of the parallax and proper
motion solutions. The stars were divided on this basis into ``probable single
stars'' and ``probably multiple stars''. We show the colour magnitude diagrams
of the two groups in figures 1 and 2.  Overlaid on the diagrams is a solar
metallicity isochrone (with an age of 11 Gyr) from Jimenez, Flynn and Kotoneva
(1998). The probable single stars are scattered around this line much less than
the multiple stars, as one would expect. 

Removing probable multiple stars reduced the sample to 449 stars (or about 2/3
of the initial sample). Despite this expedient, a small number of binaries seem
to remain in the sample (as covered in detail in section 4, Fig 10). After
removing also these suspected multiple stars, the final sample consists of 433
stars (Figure 1).

\begin{figure}
\epsfig{file=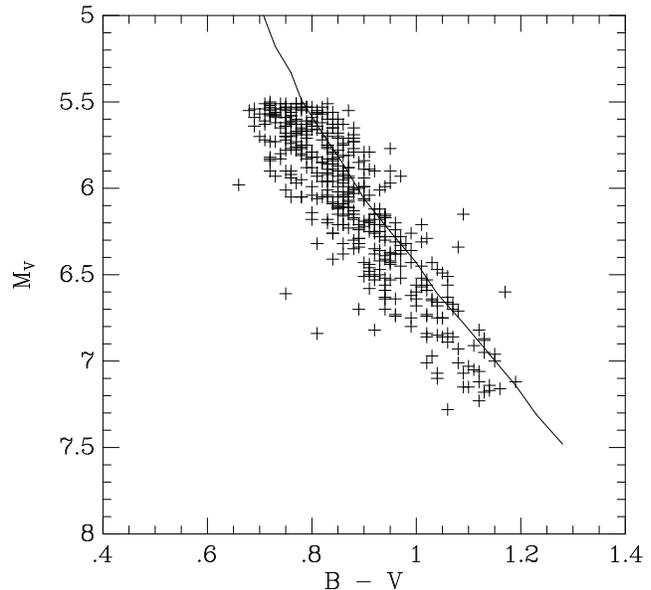,width=84mm}
\caption{The colour magnitude diagram of single stars of the data data set (433
stars).  The single stars form a clear main sequence. The line shows a solar
metallicity 11 Gyr isochrone from Jimenez, Flynn and Kotoneva, 1998).}
\end{figure}

\begin{figure}
\epsfig{file=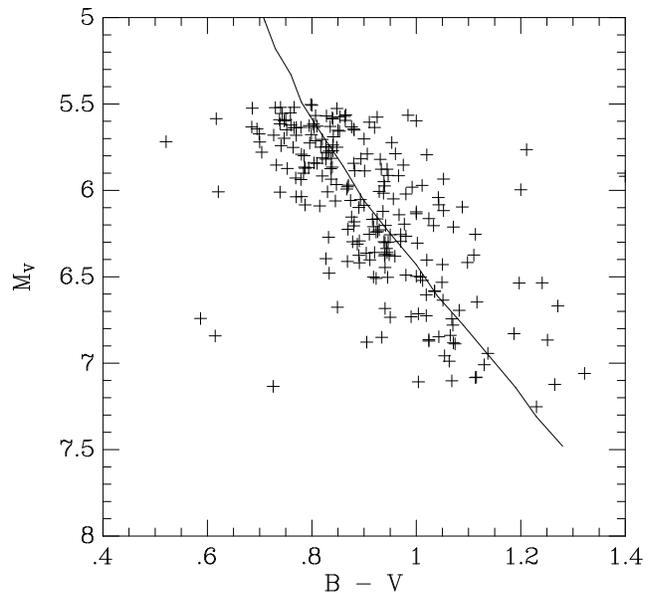,width=84mm}
\caption{The colour magnitude diagram of the multiple stars of the data set
(235 stars).  The effect of multiple stars is clear, as they tend to lie to the
right of the main sequence as tracked by the single stars.  The same 11 Gyr
isochrone is shown as in Figure 1.}
\end{figure}

\subsection{Metallicities}

Metallicities for 213 of the 433 single stars are available from the sample of
Kotoneva and Flynn (2002). The photometric metallicities are calibrated by a
sample of 34 G and K dwarfs for which accurate, spectroscopically determined
metallicity abundances, [Fe/H]$_{\mathrm spec}$ and effective temperatures
have been determined with errors of 0.05 dex and $\approx 100$ K, respectively
(Flynn and Morell, 1997). Photometric metallicities for the 213 stars come
either from a method based on Geneva $b_1$ and Johnson-Cousins $R - I$ colours
(Flynn and Morell, 1997), or from a method based on Str{\"o}mgren $m_1$ and $R
- I$ colours (Kotoneva and Flynn, 2002). Both methods give metallicities
accurate to 0.2 dex (Kotoneva and Flynn, 2002).

\section{Lower Main Sequence luminosity}

In this section we have followed the method of Jimenez, Flynn and Kotoneva
(1998, hereafter JFK) to see how well different isochrone sets fit our
sample. We have tested the following isochrone sets from the literature:

\begin{itemize}

\item{} an updated set of the MacDonald isochrones (MacDonald, private
communication; see also Jimenez and MacDonald, 1996),

\item{} the PADOVA isochrones in their newest version (Girardi et al 2000),

\item{} the Siess et al (2000) isochrones,

\item{} the Yonsei-Yale (Yi et al 2001, hereafter Y$^2$) isochrones,

\item{} the GENEVA models (Lejeune and Schaerer, 2001) were used to test both
Johnson-Cousins and the Geneva photometric systems.

\end{itemize}

In general, only a few metallicities (typically three) are computed in the
isochrone sets, meaning that only a small subset of the data can actually be
compared to the predictions of the isochrones. This was not the case for the
Yonsei-Yale (2001) isochrones, because we were able to use the code by Yi et al
(2001) to generate isochrones for a complete range of metallicities.

Table 1 shows the different metallicities and helium content of the isochrone
sets we tested. For the MacDonald set we have isochrones for ages 10, 12 and 14
Gyr. For the PADOVA, Siess, Y$^2$, and GENEVA sets we used ages 1, 5 and 10
Gyr.

\begin{table}
\begin{center}
\caption{The metallicities in $Z$ and $[{\rm Fe/H}]$ and the $Y$ value
for the isochrone sets used.}
\begin{tabular}{l|r|r|r}
\hline
Isochrone set & $Z$ & $Y$ & $[{\rm Fe/H}]$\\
 \hline
 MacDonald    & 0.004 & 0.24  &  $-$0.50  \\
 MacDonald    & 0.020 & 0.28  &  0.00     \\
 MacDonald    & 0.050 & 0.36  &  0.30     \\
              &       &       &           \\
 PADOVA       & 0.008 & 0.25  & $-$0.38   \\    
 PADOVA       & 0.019 & 0.27  & 0.00      \\    
 PADOVA       & 0.030 & 0.30  & 0.20      \\    
              &       &       &           \\
 Siess et al  & 0.010 & --    & $-$0.28   \\    
 Siess et al  & 0.020 & --    &  0.00     \\    
 Siess et al  & 0.030 & --    &  0.19     \\
              &       &       &           \\    
 Y$^2$        & 0.007 & 0.244 & $-$0.43   \\
 Y$^2$        & 0.020 & 0.270 & 0.01      \\
 Y$^2$        & 0.040 & 0.310 & 0.39      \\
              &       &       &           \\
 GENEVA       & 0.008 & 0.264 & $-$0.38   \\
 GENEVA       & 0.020 & 0.300 & 0.00      \\
 GENEVA       & 0.040 & 0.340 & 0.39      \\

\hline
\end{tabular}
\end{center}
\end{table}

\subsection{Initial testing of the isochrones}

In principle, to test an isochrone of a given metallicity, we select stars from
the sample with the same metallicity within some limit (typically $\pm 0.1$
dex) and overlay the isochrones on the colour magnitude diagram for these
stars.  In practice (as discussed in detail in section 3.2 of Jimenez, Flynn
and Kotoneva, 1998) matters are by no means so simple because of the large
scatter (0.2 dex) in the {\it observed} metallicities. Because of the shape of
the local stellar metallicity distribution, which is quite peaked, the mean
observed metallicity of stars selected from the sample will differ
substantially from their true mean metallicity. For example, for stars selected
to have an (observed) metallicity of solar, there will be many more sub-solar
metallicity stars scattered into the set than those with super-solar
metallicity.

To determine the corrections for this effect, we have carried out Monte Carlo
simulations (as in JFK) but in more detail. We proceeded as follows. The
metallicity distribution of the 433 sample stars was deconvolved (by the 0.2
dex scatter in the metallicities) to recover the true underlying distribution.
A large number of stars were then selected from this true distribution, given
a random error in the metallicity of 0.2 dex (Gaussian), and the difference
between the mean observed metallicity and the true mean metallicity computed
as a function of metallicity. The results are shown in table 2.

As an example of reading this table, consider one has selected stars with {\it
observed} metallicities in the range $0.15 < $ [Fe/H] $< 0.25$) and that they
have a mean observed metallicity of [Fe/H] $ = 0.20$. Most of these stars will
actually have scattered into this metallicity bin via observational errors from
{\it lower metallicity}, toward where the peak lies in the stellar metallicity
distribution. The simulations show that the true mean metallicity of such stars
is [Fe/H] $= 0.0$ (row 5 of the table).  Hence, in order to compare stars from
the sample to the predictions of a solar metallicity isochrone, one selects
stars with an {\it observed} metallicity of [Fe/H] $= 0.2$. Similarly, a metal
weak isochrone at [Fe/H] $= -0.5$ for example, means comparing to stars with a
mean observed metallicity of [Fe/H] $= -0.67$. Metallicities of even 0.1 dex
accuracy would alleviate these problems of comparison considerably, but
ultimately spectroscopic accuracies ($\pm 0.05$ dex) in the metallicities would
seem to be highly desirable. The present comparison of data to the predictions
of isochrones is really only a statistical one because of the relatively large
metallicity error, but the large sample size is sufficient to achieve a first
understanding of how well stellar isochrones sets fit real stars.

\begin{table}
\begin{center}
\caption{The Monte Carlo simulation of the relation between the true and
observed metallicities of stars for a scatter in the metallicity of 0.2
dex. The first column shows the true metallicity, the second the observed
metallicity and the third column the difference of the two. }
\begin{tabular}{r|r|r}
\hline
 $[{\rm Fe/H]_{true}}$ & $[{\rm Fe/H]_{obs}}$ & $ \Delta[{\rm Fe/H}] $  \\
\hline
 $-$1.28 & $-$1.38  &0.10    \\
 $-$0.68 & $-$0.83  &0.15    \\
 $-$0.50 & $-$0.67  &0.17    \\
 $-$0.38 & $-$0.48  &0.10    \\
    0.00 &    0.20  &$-$0.20 \\
    0.20 &    0.37  &$-$0.17 \\
    0.30 &    0.50  &$-$0.20 \\
\hline
\end{tabular}
\end{center}
\end{table}

\begin{figure}
\epsfig{file=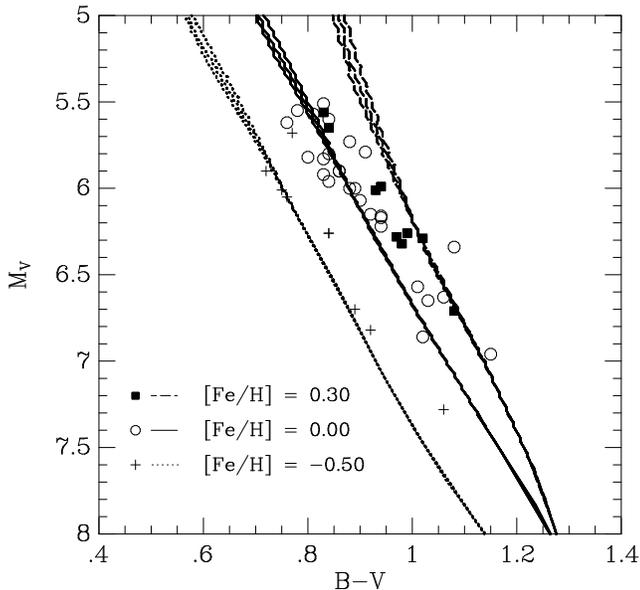,width=84mm}
\caption{Comparison of the MacDonald isochrones with appropriate stars from our
  data set.  The isochrones are from the left $[{\rm Fe/H}] = -0.50, 0.00$ and
  0.30 and shown by dotted, solid and dashed respectively. Crosses refer to
  the stars with the same metallicity as the metal-weak isochrones, open
  circles are the stars matching with the solar isochrones and squares are the
  stars matching with the metal rich isochrones. The metal weak and solar
  isochrones fit the data very well, while the metal rich isochrone does not
  provide such a good fit. The ages of the isochrones are 10, 12 and 14 Gyr.}
\end{figure}

We now move on to comparing the available data with the isochrones.

\subsection{MacDonald isochrones} 

The MacDonald isochrones are available for metallicities of [Fe/H] $= -0.50,
0.0, 0.30$ (MacDonald, private communication; see also Jimenez and MacDonald,
1996). They have been transformed into the observational plane by use of the
Kurucz atmospheres models (Kurucz 1993). We have selected stars from the
sample with observed metallicities of $-0.67\pm0.1, 0.20\pm0.1$ and
$0.50\pm0.1$ for comparison with these isochrones (the selected metallicities
differ from the isochrone metallicities because of the effects of
observational scatter, as explained in the previous section). Fig.~3 shows the
MacDonald isochrones overlaid on the colour magnitude diagram of these three
sets of stars from the sample. All isochrone sets in what follows are shown
for three different ages, although age of course has very little effect for
the stars of the spectral type being considered (K dwarfs). Solar metallicity
stars are shown by open circles and the isochrones (for three ages) by solid
lines. Sub-solar metallicity stars are shown by crosses and the sub-solar
isochrones by dotted lines, and the super-solar metallicity stars by filled
squares and the super-solar metallicity isochrones by dashed lines. Stars and
isochrones are denoted in this system throughout the comparison (i.e. Figures
3, 4, 5, 6, 8 and 9).

The MacDonald solar metallicity isochrones in figure 3 match the data quite
well, and are excellent for the low metallicity case.  Although the super-solar
isochrones achieve good agreement with the data for $M_V > 6$, stars more
luminous than this appear to be hotter than the models (by about 200 K).
Before we discuss this disagreement in more detail it is worth noting that the
MacDonald isochrones use $\Delta Y/\Delta Z = 2.5$, for the metallicity range
considered (see also table 2). As it has been pointed out by several
researchers (see e.g. Pagel and Portinari, 1998 and references therein) the
width of the main sequence at a fixed luminosity for dwarfs also depends on
the value of $\Delta Y/\Delta Z$ although a similar effect is also due to
variations with metallicity in the mixing length parameter (e.g. Jimenez and
MacDonald, 1996). We have experimented with changing $\Delta Y/\Delta Z$ by
$\pm 0.5$.  This experiment is qualitative only, because our starting point is
isochrones which are not yet a good fit to the super-solar metallicity stars.
Adopting values of $\Delta Y/ \Delta Z$ in this range leads to no significant
improvements in the overall fits of the isochrones. For example, increasing
$\Delta Y/ \Delta Z$ to 3 (a very modest increase) helps to fit the luminous
metal rich stars but the fit to the low luminosity metal-rich dwarfs
deteriorates. An alternative explanation is that the mixing length parameter
might change with luminosity (increase) for metal rich stars.

\subsection{PADOVA isochrones} 

Fig.~4 shows a similar comparison as in Fig.~3 but this time using the PADOVA
isochrones. The first thing to note is that the solar isochrone provides a fit
of nearly the same quality as the MacDonald isochrones. On the other hand both
the low and high metallicity isochrones do not provide adequate fits to the
data. The metal-rich isochrone does provide a good fit to the luminous stars
($M_V < 6$) but fails to provide a good fit to the less luminous dwarfs. This
can be traced back to the large value of $\Delta Y/ \Delta Z$ used (2.7 for
$Z=0.03$) in these isochrones (we found the same effect for the MacDonald
isochrones). Furthermore, for the low metallicity isochrone the adopted $\Delta
Y/ \Delta Z$ value is rather low (1.8 for $Z=0.008$), and may be the reason for
the isochrone to be sub-luminous relative to the data.

\begin{figure}
\epsfig{file=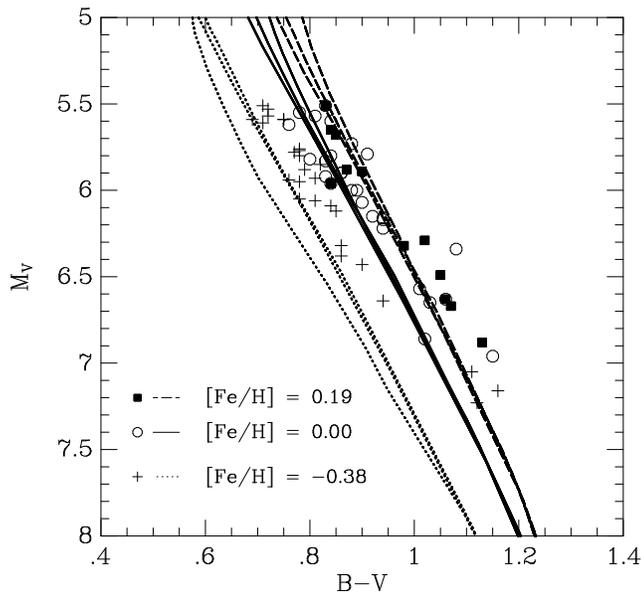,width=84mm}
\caption{Comparison of the PADOVA isochrones (Girardi et al, 2001) with
appropriate stars from our data set. The isochrones are from the left $[{\rm
Fe/H}] = -0.38, 0.00$ and 0.19 and shown by dotted, solid and dashed
respectively. Crosses refer to the stars with the same metallicity as the
metal-weak isochrones, open circles are the stars matching with the solar
isochrones and squares are the stars matching with the metal rich
isochrones. None of the isochrones fit particularly well. Three ages are used
for the isochrones, 1, 5 and 10 Gyr.}
\end{figure}

\subsection{Siess et al isochrones} 

In Fig 5 the Siess et al (Siess, Dufour and Forestini, 2000) isochrones are
shown compared to the data.  The slope of these isochrones matches the data
well and the metal weak isochrones are correctly located, but the solar and the
metal rich isochrones do not match the data as well, being systematically too
faint. Especially interesting is the solar composition metallicity isochrone
which appears to be too faint by a few tenths of a magnitude, but given the
small number of stars this difference may not be significant (i.e. the error in
these comparisons is dominated by systematic effects which are plausibly at
least a few $\times$ 0.1 mag). The disagreement is more severe for the metal
rich case.

\begin{figure}
\epsfig{file=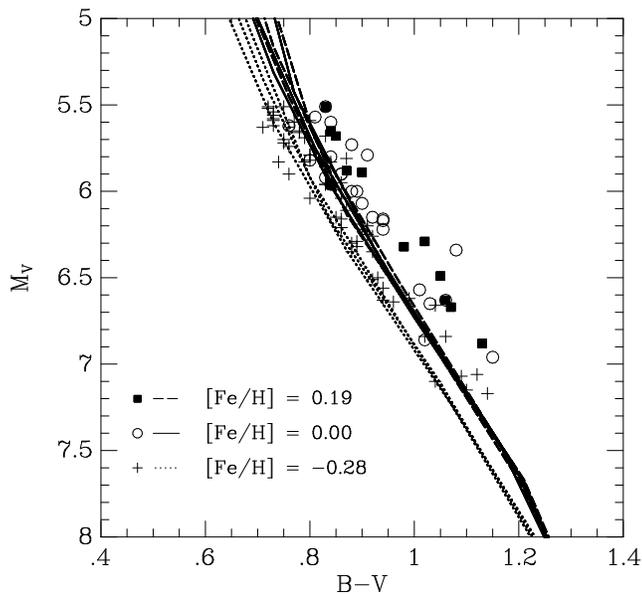,width=84mm}
\caption{Comparison of the Siess et al (2000) isochrones with appropriate stars
from our data set.  The isochrones are from the left $[{\rm Fe/H}] = -0.28,
0.0$ and 0.19 and dotted, solid and dashed respectively.  The symbols are the
same as in previous figures. The ages are the same as for the PADOVA
isochrones, i.e. 1, 5 and 10 Gyr.}
\end{figure}

\subsection{Y$^2$ isochrones} 

The Y$^2$ (i.e. Yonsei-Yale) isochrones are compared in Fig.~6 with the
Hipparcos data, choosing the Lejeune et al. (1998) library to transform from
effective temperature and luminosity to colour and absolute magnitude.  The
solar metallicity isochrone provides a good fit to the data and the overall
slope matches the data well. On the other hand, the low and high metallicity
isochrones do not match the data. For the $Y^2$ isochrones $\Delta Y / \Delta
Z = 2$ for the whole metallicity range. It would be very interesting to have
these isochrones recomputed for values of, say, $\Delta Y / \Delta Z = 3$ and
4, as higher values might alleviate the problem with fitting the higher and
lower metallicity isochrones.

\begin{figure}
\epsfig{file=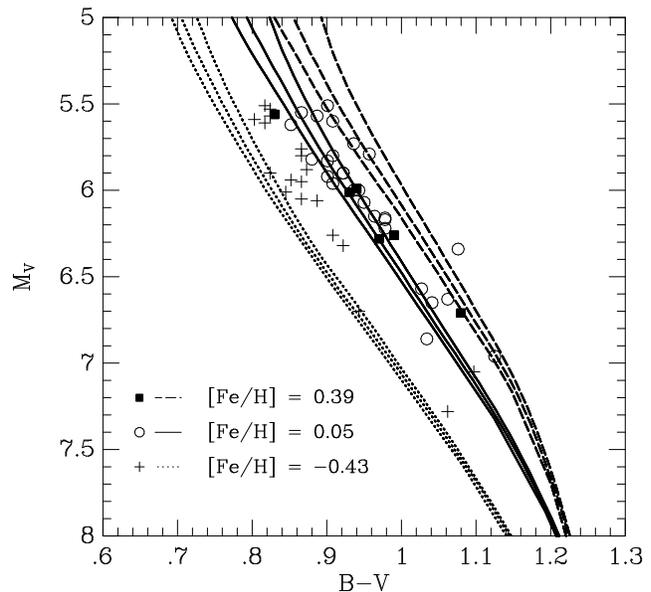,width=84mm}
\caption{Comparison of the Y$^2$ isochrones with the Hipparcos data. The
isochrones are from the left $[{\rm Fe/H}] = -0.43, 0.01$ and 0.39 and are
dotted, solid and dashed lines respectively.  The symbols are the same as in
previous figures. The ages are the same as for PADOVA isochrone set, i.e. 1, 5
and 10 Gyr.}
\end{figure}

A major shortcoming with our comparison technique is that we are restricted to
using only stars at the appropriate metallicity for the isochrones being tested
(so that most of the data is not being used). This problem can be overcome in
the case of the Y$^2$ isochrones because interpolation routines for
metallicities between the isochrones are available (Yi et al, 2001).  Using
their code we created a continuous set of isochrones for metallicities between
$-1.45 < $ [Fe/H] $ < 0.85$ in steps of 0.1 dex.  Each isochrone was then
compared to those stars for which the observed metallicity matched the
isochrone metallicity with 0.05 dex (and taking into account the systematic
shift between mean observed metallicity and mean true metallicity described in
section 3.1). This makes use of the whole data set.

The results are shown in Fig.~7. We plot the difference between the Hipparcos
measured absolute magnitude and the magnitude of the isochrone at the star's
colour, $\Delta(Y^2)$, as a function of the metallicity, [Fe/H]. If the
isochrones match the data, then $\Delta(Y^2) = 0$ (horizontal line). There is
clearly a systematic shift between the isochrones and the data, in the sense
that the isochrones are a little too faint. Any trend with metallicity, which
might indicate a too low value of $\Delta Y / \Delta Z$ or incorrect high $Z$
opacity in the stellar models, cannot be concluded with confidence from these
data.  Nevertheless, $\Delta Y / \Delta Z = 2$ may be a too low value; the
value adopted for the MacDonald isochrones (2.5) would push the Y$^2$
isochrones into better agreement with the data.

\begin{figure}
\epsfig{file=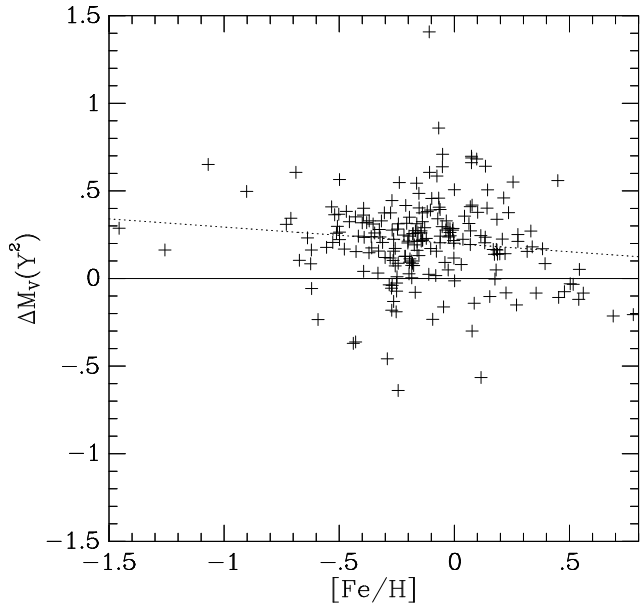,width=84mm}
\caption{The relation between $\Delta M_V$ (stellar absolute magnitude relative
to predicted absolute magnitude for the $Y^2$ isochrones) as a function of
metallicity. The crosses are the stars from the Hipparcos sample. Perfect
agreement between the isochrones and the data would be obtained for $\Delta M_V
= 0$ (solid line). The dashed line shows a least squares fit to the
data. Although there is a slight trend with metallicity, which might indicate
that the isochrone fit deteriorates slightly going toward lower metallicity,
the trend is not statistically significant. In general, the Y$^2$ isochrones
stand up well to the comparison with the available data.}
\end{figure}

\subsection{GENEVA isochrones} 

The last group of isochrones we studied is the GENEVA sets (Lejeune and
Schaerer, 2001). We compared these isochrones to the data for colour magnitude
diagrams in both the $B - V$ and $b_1$ colours (since the GENEVA isochrones
predict $b_1$ colour and this colour was used to measure the metallicities for
about half the sample). The comparisons are shown in Fig 8. The isochrones
appear to be generally under-luminous at a given colour in $B-V$. Interestingly,
the isochrones fit the $b_1$ colours better than they do the $B-V$ colours.

\begin{figure}
\epsfig{file=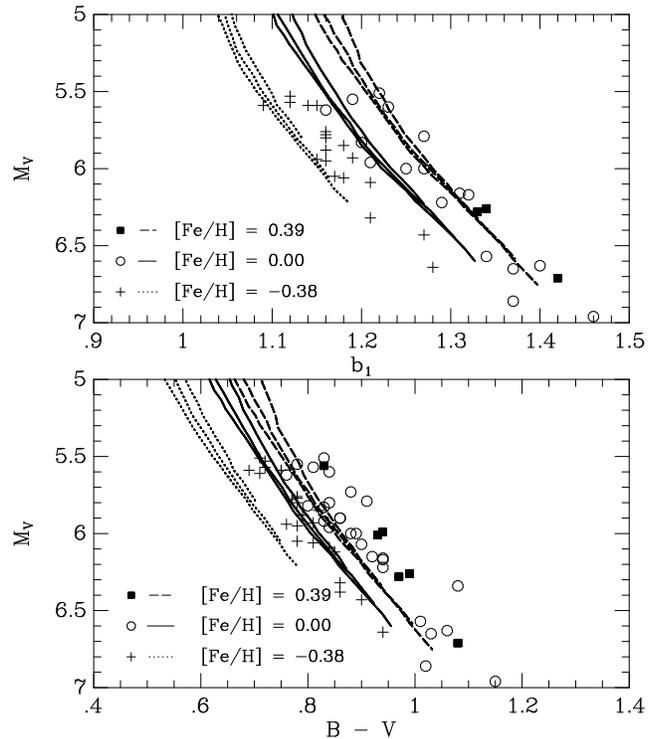,width=84mm}
\caption{Comparison of the GENEVA isochrones with stars from the Hipparcos
sample at the appropriate metallicity. The upper panel shows the Geneva
isochrones in the $M_V$ versus $b_1$ plane while the lower panel shows $M_V$
versus $B - V$ colour. The dotted lines in both panels show metallicity $[{\rm
Fe/H}] = -0.38$, the solid lines the solar metallicity and the dashed lines
refer to the metal-rich isochrone, $[{\rm Fe/H}] = 0.39$.  Crosses refer to the
stars with the same metallicity as the metal-weak isochrones, open circles are
the stars matching with the solar isochrones and squares are the stars matching
with the metal rich isochrones.}
\end{figure}

\subsection{Summary of isochrone comparisons}

We have tested several sets of isochrones available in the literature.  The
MacDonald set provides an excellent fit for solar and sub-solar metallicities,
but it is not as successful with the super solar models. The PADOVA set provide
a good fit to the solar observations but are less satisfactory for metal-poor
and rich stars. For the PADOVA isochrones this may be due to the low adopted
value of $\Delta Y/ \Delta Z = 1.7$ for low metallicity and the high value
(2.7) used for metal rich isochrones. In the case of the Y$^2$ isochrones the
mismatch at other metallicities than solar is due to the low value (2) adopted
for $\Delta Y/ \Delta Z$. The Siess et al isochrones fit well the metal weak
stars but fail to fit the solar and solar-rich stars.  The Y$^2$ and the
GENEVA isochrones have the right slope in the colour magnitude diagram, and
the Y$^2$ fit with the solar metallicity stars is quite good. Both Y$^2$ and
GENEVA sets produce metal weak or rich isochrones that are too faint (or too
hot), and the GENEVA system has difficulties to fit even the solar
observations.

\section{Empirical calibrations of K dwarf luminosity}

The comparisons of the isochrones in the previous section usually only made use
of stars of metallicity within $\pm$ 0.1 of the isochrone metallicity. Even
this simple test already shows that none of above models fits the metal-rich
isochrones.

In Fig 9 we show the old JFK isochrones. As seen in the plot the solar
metallicity and the metal weak stars follow the isochrones well; both the slope
and the position appear to be satisfactory.

\begin{figure}
\epsfig{file=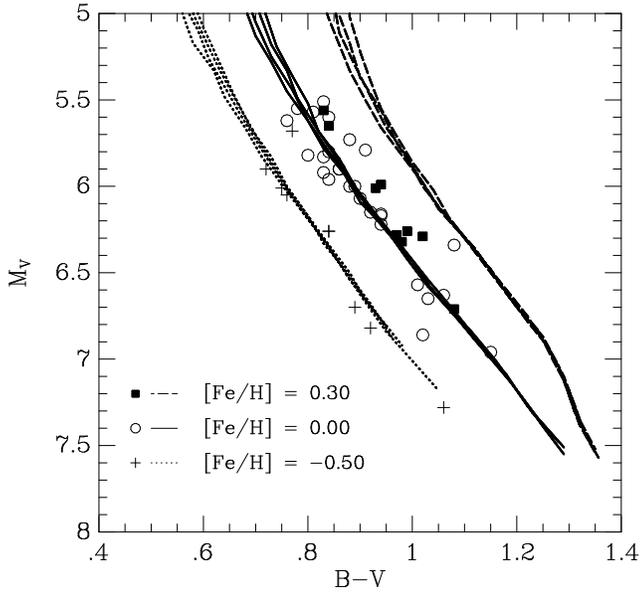,width=84mm}
\caption{Our single stars over-plotted by the JFK isochrones. The lines are from
the left $[{\rm Fe/H}] = -0.5, 0.0$ and 0.3 and dotted, solid and dashed
respectively. Crosses refer to the stars with the same metallicity than the
metal-weak isochrone, open circles are the stars matching with the solar
isochrone and squares are the stars referring to the metal rich isochrone. The
metal weak isochrone fits the data quite well and the solar one fits the data
excellently. Obviously some more work has to be done with the metal rich
isochrone. The ages of the isochrones are 8, 11, 13 and 15 Gyr.}
\end{figure}

We therefore decided to make an empirical calibration of the main sequence
luminosity as a function of the metallicity which can be used to test any
isochrone. We proceeded as follows. The JFK solar isochrones (hereafter JFK0),
are in excellent agreement with the data as seen in Fig 9.  The JFK0 with age
11 Gyr was chosen as a fiducial line. The luminosity $M_V$ of this isochrone
as a function of $B - V$ and $R - I$ is shown Table 3.

\begin{table}
\begin{center}
\caption{Luminosity $M_V$ of the JFK solar metallicity isochrone 
as a function of $B - V$ and $R - I$.}
\begin{tabular}{r|r|r|r|r|r}
\hline
$M_V$ & $B - V$ & $ R - I$  & $M_V$ & $B - V$ & $ R - I$  \\
\hline
7.48  & 1.28 & 0.73 &5.33  & 0.76 & 0.40\\
7.31  & 1.23 & 0.69 &5.18  & 0.73 & 0.39\\
7.14  & 1.19 & 0.64 &5.02  & 0.71 & 0.38\\
6.96  & 1.14 & 0.61 &4.85  & 0.69 & 0.37\\
6.78  & 1.09 & 0.58 &4.69  & 0.68 & 0.37\\
6.61  & 1.04 & 0.55 &4.65  & 0.68 & 0.37\\
6.43  & 1.00 & 0.52 &4.50  & 0.67 & 0.37\\
6.25  & 0.95 & 0.51 &4.34  & 0.68 & 0.37\\
6.06  & 0.90 & 0.48 &4.19  & 0.70 & 0.38\\
5.86  & 0.86 & 0.45 &4.03  & 0.77 & 0.41\\
5.68  & 0.82 & 0.43 &4.00  & 0.86 & 0.44\\
5.49  & 0.78 & 0.40 &      &      &     \\
\hline
\end{tabular}
\end{center}
\end{table}

\begin{figure}
\epsfig{file=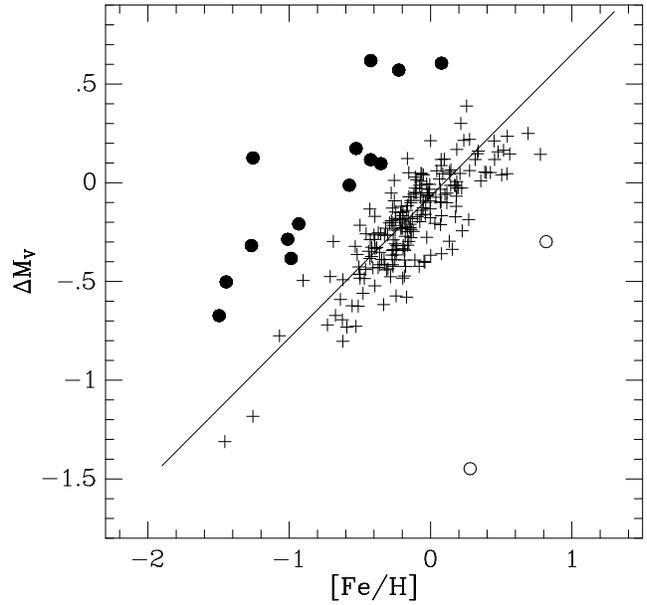,width=84mm}
\caption{The difference $\Delta M_V$ between stellar absolute magnitude (as
measured by Hipparcos) and a fiducial solar metallicity isochrone (JFK0) at the
colour of the star, plotted as a function of metallicity. Most of the stars
(crosses) mark out a well delineated sequence, which we interpret as a
luminosity/metallicity relation for the main sequence.  The stars marked by
filled circles are outliers, which we regard as probably unrecognized
multiples. Two open circles indicate stars which probably have incorrect
colours, and have been disregarded.}
\label{mvdifvsfeh}
\end{figure}

\begin{figure}
\epsfig{file=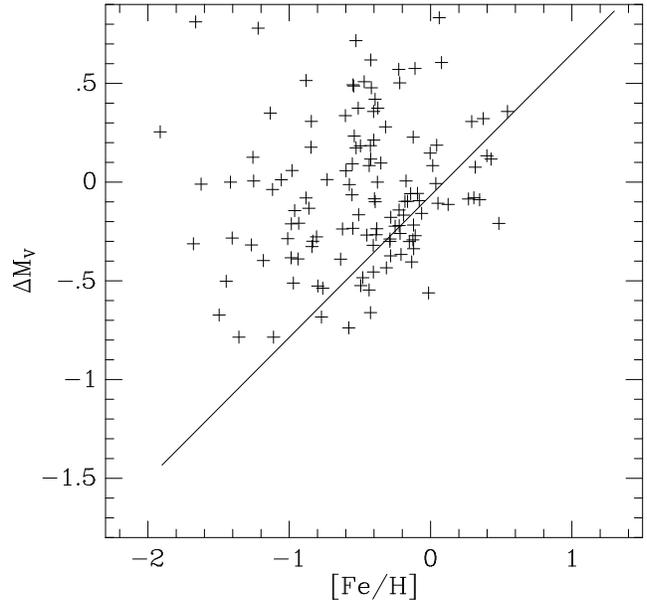,width=84mm}
\caption{The same quantity, $\Delta M_V$, as in figure \ref{mvdifvsfeh}, but
for the stars identified as multiples by Hipparcos, as a function of
metallicity. For these stars the metallicity estimate is likely incorrect
because of the contaminating light of the companion(s). The multiples are much
more scattered than the single stars (figure \ref{mvdifvsfeh}), and also tend
to be up to $\approx 0.75$ mag more luminous, as one would expect for
multiples. Many of these multiples lie in the region where the suspected
multiples are flagged (by filled circles) in figure \ref{mvdifvsfeh}.}
\label{mvdifvsfehbina}
\end{figure}

For all stars with known metallicity we measure the star's distance from the
JFK0 isochrone, $\Delta M_V$ (i.e. relatively to the fiducial line at that
colour). The results are shown in Fig 10. For most of the stars there is a
simple relation between the metallicity and luminosity (relative to
JFK0). However, there are clearly some outliers, and these have been marked as
solid circles. We identify these as remaining {\it unrecognized multiple stars}
in the basic sample. We cannot prove that they are not just outliers, but we
can assemble circumstantial evidence for this assertion. In Fig 11 we plot
$\Delta M_V$ vs. $[{\rm Fe/H}]$ for the known multiple stars (identified as
such by Hipparcos).\footnote{There were two stars well below the
metallicity-luminosity relation (shown as open circles in Fig 10). Plausibly
their colours are simply incorrect, and they have not been used further} In
this figure, multiple systems are seen to be brighter than their single
counterparts at the same metallicity by approximately 0.75 magnitude, precisely
as one would expect. Further evidence for this assertion is shown in Fig 12,
where we plot the colour-colour relation for the single stars using the same
symbols as in Fig 10. The suspected multiple stars tend to lie above the
sequence defined by the rest of the stars, and this is also where the known
multiple stars lie when plotted in this plane. The 14 suspected multiples
flagged in figure 10, would represent approximately a further 5\% in multiple
systems in addition to the 30\% removed on the basis of Hipparcos. We suspect
that the stars marked by filled circles in Figs 10 and 12 are unrecognized
multiple stars, but this would require further study to prove.

\begin{figure}
\epsfig{file=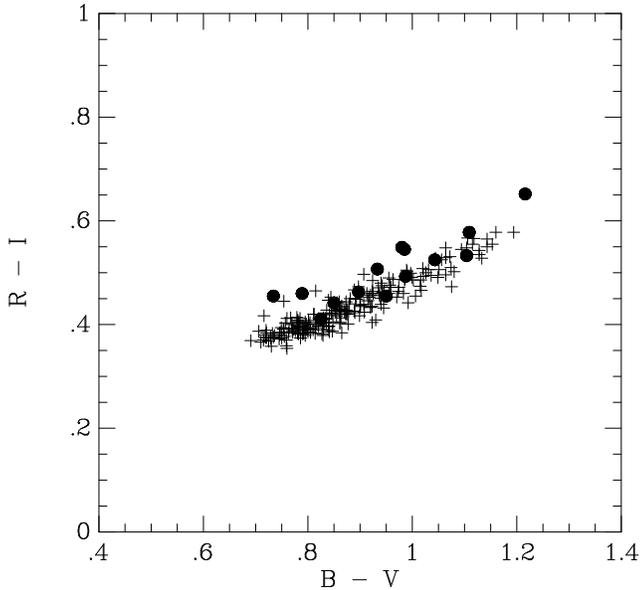,width=84mm}
\caption{$B - V$ plotted against $R - I$ for the sample stars. The crosses show
  the single stars and the filled circles the ``suspected multiple stars'',
  (see figures \ref{mvdifvsfeh} and \ref{mvdifvsfehbina}).} 
\end{figure}

\begin{figure}
\epsfig{file=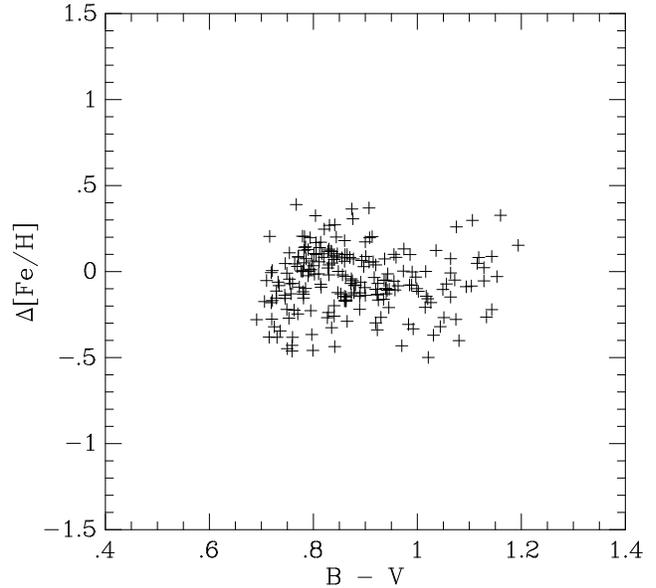,width=84mm}
\caption{The difference between metallicities derived from the luminosity of a
star relative to the solar metallicity isochrone versus photometrically
obtained metallicities, as a function of stellar colour.  No apparent residuals
remain in the difference as a function of colour, indicating there is no
systematic error in the fiducial isochrone as a function of colour
(i.e. temperature). The scatter is approximately 0.2 dex.}
\end{figure}

\subsection{The metallicity-luminosity relation for K dwarfs}

From Fig 10 it is clear that the luminosity of the bulk of the stars
(i.e. those marked by crosses) relative to JFK0 correlates well with the
metallicity. We have fitted this correlation with the following relation (shown
by a solid line in Fig 10):

\begin{equation}
\Delta M_V = 0.84375 \times {\rm [Fe/H]} - 0.04577, 
\end{equation}

which can be inverted to give a metallicity index, [Fe/H]$_{\rm KF}$:

\begin{equation}
{\rm [Fe/H]_{KF}} = 1.185 \times \Delta M_V + 0.054.
\end{equation}

To test that this relation is independent of the colour of the star we plot the
difference between the photometrically defined abundance and the metallicity
based on the luminosity, $\Delta [{\rm Fe/H}] = [{\rm Fe/H}]_{\rm KF} - [{\rm
Fe/H}]$, vs. the $B - V$ colour. As can be seen in Fig 13, the difference is
independent of colour, indicating no residual temperature effects are present
in the calibration.

Equation (1) was used to produce isochrones of various metallicity, simply
offset from JFK0. In Fig 14 is shown the JFK0 11 Gyr isochrone and the
artificial isochrones with $[{\rm Fe/H}]$ = $-$1.8, $-$1.5, $-$1.0, $-$0.8,
$-$0.6, $-$0.4, 0.2, 0.0, 0.2, 0.4 and 0.6 over-plotted with our single stars.
Most of the nearby stars lie between $[{\rm Fe/H}]$ = $-$0.6 and 0.2 with very
few metal weak stars. There are three apparent sub-dwarfs which happen to fit
the $[{\rm Fe/H}]$ = $-$1.8 isochrone: one is certainly a bone fide halo
subdwarf (HD103095) because it has high space velocities; the other two have
modest velocities and are probably not subdwarfs (HD 120559 and HD 145417).

\begin{figure}
\epsfig{file=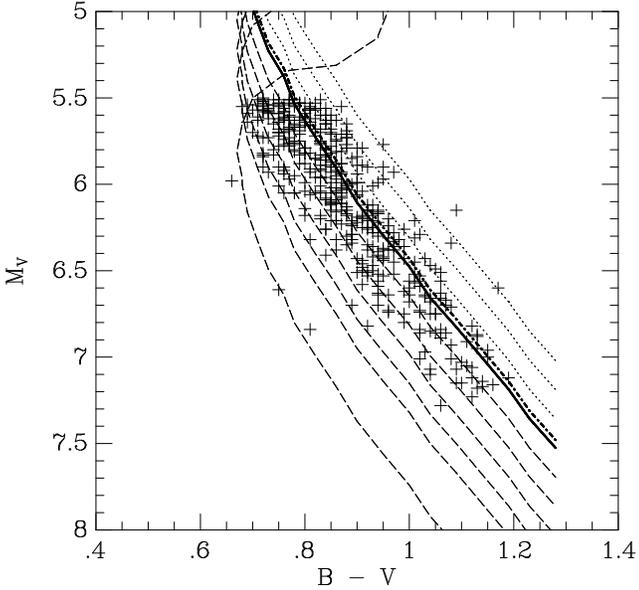,width=84mm}
\caption{Isochrones as a function of metallicity, relative to the fiducial
  solar metallicity isochrone JFK0 computed via Eq.~1. The solid line is the
  JFK0 11 Gyr isochrone. The dotted-short dashed line is the solar isochrone
  calculated on the basis of Eq.~1. The dotted lines above these lines are
  isochrones for [Fe/H] $ = 0.2, 0.4$ and 0.6 from left to right and the
  dashed lines below solar isochrones are for [Fe/H] $ = -1.8$, $-1.5$, $-1.0,
  -0.8, -0.6, -0.4$ and $-0.2$ also from left to right.}
\end{figure}

\section{Accuracy of the metallicity-luminosity relation}

The photometric metallicities used to derive the luminosity relation (Eq.~1)
are ultimately based on spectroscopically obtained metallicities for G and K
dwarfs.  All of these dwarfs have accurate metallicities, colours and
parallaxes, and we can check that they lie on the relation (Eq.~1). This is a
self-consistency check on the photometric metallicity calibration. We decided
as a simple exercise to check the self-consistency of the method and yielded a
remarkable result: the position of a K dwarf within the main sequence on a
colour-magnitude diagram is {\it very} tightly correlated with metallicity.

The spectroscopic sample of 15 stars with which we made the check is shown in
Table 4. Most of the stars come from Flynn and Morell (1997), and a
super-solar metallicity K dwarf has been taken from Feltzing and Gonzalez
(2001). All the stars are ``K dwarfs'', i.e. in the absolute magnitude range
$5.5 < M_V < 7.3$. The table shows the HD number, $B-V$ colour, absolute
magnitude $M_V$ (from the Hipparcos parallax) and the difference between the
absolute magnitude of the star and the absolute magnitude of the JFK0
isochrone (at the colour of the star).  We then show the spectroscopically
determined metallicity [Fe/H]$_{\mathrm spec}$ and the metallicity
[Fe/H]$_{\mathrm KF}$, derived from Eq.~1.

\begin{table}
\begin{center}
\caption{K dwarfs for which spectroscopically determined metallicities are
available from Flynn and Morell (1997) and Feltzing and Gonzalez
(2001). Columns are: HD number, $B-V$ colour and $V$-band absolute magnitude,
$M_V$; the difference between the star's absolute magnitude and the absolute
magnitude of the JFK0 isochrone (see figure \ref{CMDkdwarfs}, panel (a)); the
spectroscopic metallicity, [Fe/H]$_{\mathrm spec}$, and the metallicity derived
from Eq.~1 [Fe/H]$_{\mathrm KF}$.}
\begin{tabular}{rrrrrr}
\hline
HD    &  $B-V$  & $M_V$&$\Delta(M_V)$ & [Fe/H]$_{\mathrm spec}$ & [Fe/H]$_{\mathrm KF}$ \\
\hline
      4628  &  0.89  &  6.38 &$ -0.37 $ & $   -0.40$ & $   -0.38 $ \\
     10700  &  0.73  &  5.68 &$ -0.52 $ & $   -0.50$ & $   -0.57 $ \\
     13445  &  0.81  &  5.93 &$ -0.29 $ & $   -0.28$ & $   -0.29 $ \\
     25329  &  0.86  &  7.18 &$ -1.30 $ & $   -1.70$ & $   -1.49 $ \\
     26965  &  0.82  &  5.92 &$ -0.24 $ & $   -0.30$ & $   -0.23 $ \\
     32147  &  1.05  &  6.49 &$  0.15 $ & $    0.28$ & $    0.23 $ \\
     72673  &  0.78  &  5.95 &$ -0.46 $ & $   -0.35$ & $   -0.49 $ \\
    100623  &  0.81  &  6.06 &$ -0.43 $ & $   -0.25$ & $   -0.45 $ \\
    103095  &  0.75  &  6.61 &$ -1.31 $ & $   -1.40$ & $   -1.50 $ \\
    134439  &  0.77  &  6.74 &$ -1.33 $ & $   -1.57$ & $   -1.52 $ \\
    134440  &  0.85  &  7.08 &$ -1.26 $ & $   -1.52$ & $   -1.44 $ \\
    149661  &  0.83  &  5.82 &$ -0.11 $ & $    0.00$ & $   -0.07 $ \\
    192310  &  0.88  &  6.00 &$ -0.05 $ & $   -0.05$ & $   -0.01 $ \\
    209100  &  1.06  &  6.89 &$ -0.23 $ & $   -0.10$ & $   -0.22 $ \\
    216803  &  1.09  &  7.07 &$ -0.27 $ & $   -0.20$ & $   -0.27 $ \\
\hline
\end{tabular}
\end{center}
\end{table}

\begin{figure}
\epsfig{file=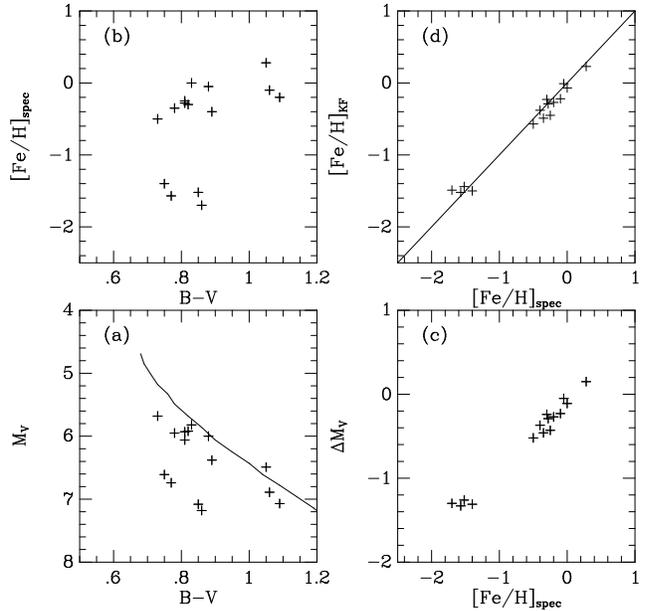,width=84mm}
\caption{Metallicity-luminosity relation for K dwarfs. Panel (a): colour
magnitude diagram for K dwarfs from table 4, for which accurate spectroscopic
metallicities are available. The solid line shows the JFK0, solar metallicity
isochrone. Panel (b): Metallicity versus $B-V$ colour for the K dwarfs. Panel
(c): Difference $\Delta(M_V)$ between the star's absolute magnitude $M_V$ and
the JFK0 isochrone at the stars $B-V$ colour. Panel (d): metallicity
[Fe/H]$_{\mathrm KF}$ derived using $\Delta(M_V)$ in Eq.~1, versus the
spectroscopic metallicity, [Fe/H]$_{\mathrm spec}$. The line shows a 1:1
relation.  The scatter in the relation is less than 0.1 dex.}
\label{CMDkdwarfs}
\end{figure}

A colour magnitude diagram ($B-V$ versus $M_V$) for the stars in table 4 is
shown in panel (a) of figure \ref{CMDkdwarfs}, along with the JFK0
isochrone. Panel (b) shows the metallicities of the stars as a function of
$B-V$ colour. Panel (c) shows the absolute magnitude difference $\Delta(M_V)$
between the star's $M_V$ and the absolute magnitude of the JFK0 isochrone (at
the star's $B-V$ colour), as a function of the spectroscopically determined
metallicity, [Fe/H]$_{\mathrm spec}$. There is a {\it very} tight relationship
between $\Delta(M_V)$ and [Fe/H]$_{\mathrm spec}$ (the scatter in a least
squares fit is a mere 0.08 magnitudes). Clearly, luminosity on the main
sequence, relative to the solar isochrone, is very tightly correlated with
metallicity. Panel (d) shows the relationship between spectroscopic metallicity
[Fe/H]$_{\mathrm spec}$ and the metallicity computed via Eq.~1,
[Fe/H]$_{\mathrm KF}$. The line shows a 1:1 relation. Clearly, Eq.~1 produces a
a very accurate metallicity estimate for the stars: the scatter in the
residuals is a mere 0.08 dex in [Fe/H]. Since the spectroscopic metallicities
are accurate to approximately 0.05 dex, the metallicity [Fe/H]$_{\mathrm KF}$
(derived directly from the position of the star relative to the JFK0 isochrone
in the CMD) appears to be as accurate as a spectroscopically measured
metallicity. Eq.~1 turned out to yield a far more accurate metallicity than we
had anticipated.

\subsection{Comparison with open and globular cluster K dwarf luminosities}

We have performed a preliminary check on the metallicity-luminosity relation by
using it to compute expected apparent magnitudes of K dwarfs in some well
studied open and globular clusters, for which sufficiently deep $BV$ photometry
exists. A suitable colour to compare the observed main sequence with the JFK0
isochrone is at $B-V = 0.9.$ Let the main sequence magnitude at this colour be
denoted $M_V(0.9)$. For solar metallicity, $M_V(0.9) = 6.17$ (from Eq.~1 and
Table 3).

\begin{table*}
\begin{center}
\caption{Check on the expected absolute magnitudes for open and globular
cluster K dwarfs at $(B-V)_0 = 0.9$ using the metallicity-luminosity relation
(i.e. Eq.~1). Columns are the name; the adopted metallicity [Fe/H] and
reddening $E(B-V)$; the $V$ band true distance modulus (which is as far as
possible independent of a main sequence fitting); the absolute magnitude
$M_V(0.9)$ expected from Eq.~1 at $(B-V)_0 = 0.9$ and the observed absolute
magnitude $M_V$}
\begin{tabular}{lrrrrrrl}
\hline
Name     & [Fe/H]  &$E(B-V)$& $(m-M)_0$ & $V$   & $M_V(0.9)$  &$M_V$  & Colour magnitude diagram    \\ 
Hyades   &  0.15   & 0.00~~~&  1.33~~~  &  7.3~ & 5.98 ~~~    & 6.1~  & Haywood (2001)              \\
M67      & $-0.1~$ & 0.05~~~&  9.58~~~  & 15.9~ & 6.19 ~~~    & 6.2~  & Montgomery et al (1993)     \\ 
NGC 2420 & $-0.4~$ & 0.05~~~& 11.95~~~  & 18.6~ & 6.44 ~~~    & 6.4~  & Anthony-Twarog et al (1990) \\ 
47 Tuc   & $-0.7~$ & 0.05~~~& 13.1:~~~  & 20.1~ & 6.73 ~~~    & 7.0:  & Hesser et al (1987)         \\ 
M4       & $-1.2~$ & 0.37~~~& 11.4:~~~  & 19.8~ & 7.12 ~~~    & 7.3:  & Alcaino et al (1997)        \\ 
NGC 6752 & $-1.6~$ & 0.04~~~& 13.05~~~  & 20.6: & 7.46 ~~~    & 7.4:  & Penny and Dickens (1997)    \\ 
\hline
\end{tabular}
\end{center}
\end{table*}

\begin{itemize}

\item For the Hyades, we compute an absolute magnitude of the main sequence at
$B-V = 0.90$ of $M_V(0.9) = 5.98$, which is in good agreement with the observed
value of $M_V = 6.1$ (Haywood 2001).

\item For the open cluster M67 ([Fe/H]$ = -0.1$, $E(B-V) = 0.05$ and $(m-M)_0 =
9.6$ (Richer et al 1998, Montgomery et al 1993). Very deep CCD photometry by
Montgomery et al (1993) shows that the apparent magnitude of K dwarfs at $B-V =
0.95$ is $V \approx 15.8$. This leads to an observed absolute magnitude at
$(B-V)_0 = 0.9$ of $M_V \approx 6.2$ compared to an expected absolute magnitude
of $M_V(0.9) = 6.19$.

\item The open cluster NGC 2420 has been observed with HST by von Hippel and
Gilmore (2000), in order to detect the white dwarf cooling sequence. The
advantage here is that the distance modulus is not determined from the main
sequence. For their adopted metallicity of [Fe/H] $= -0.4$, reddening of
$E(B-V) = 0.04$ and true distance modulus of 12.10, Eq.~1 gives an absolute
magnitude at $(B-V)_0 = 0.9$ of $M_V(0.9) = 6.44$, close to the observed value
of $M_V = 6.4$ (from an apparent $V \approx 18.6$, read from figure 7 of
Anthony-Twarog et al 1990).

\item The globular cluster 47 Tuc has a metallicity of [Fe/H] $\approx -0.7$,
$E(B-V)=0.05$ and a (white dwarf based) true distance modulus has been
determined by Zoccalli et al (2001) of 13.09: the apparent magnitude of the
main sequence at $B-V = 0.95$ is $V \approx 20.3$, or $M_V = 7.0$.  This is not
in good agreement with the absolute magnitude $M_V(0.9) = 6.73$ expected from
Eq.~1. However, Zoccalli et al (2001) found a distance modulus for 47 Tuc
somewhat shorter than obtained by traditional techniques; in particular, the
horizontal branch based distance modulus is $\approx 13.4$ (corresponding to a
main sequence absolute magnitude $M_V = 6.8$) which does improve the agreement.
Of all the clusters studied here with distance moduli based on white dwarfs,
47 Tuc is the only one in significant disagreement with the distance modulus
using traditional techniques (horizontal branch and/or main sequence
fitting). 

\item The globular cluster M4 has [Fe/H] $ = -1.2$ and $E(B-V) = 0.37$. The
expected luminosity at $(B-V)_0 = 0.9$ is $M_V(0.9) = 7.12$. Richer et al
(1997) have derived an apparent $V$ band distance modulus for M4 of 12.5 from a
fit to the white dwarf sequence on very deep HST images.  Reading from the
colour magnitude diagram in Alcaino et al (1997), the apparent magnitude at
$B-V = 1.27$ is $V \approx 19.8$, which yields an absolute magnitude of $M_V =
7.3$ in good agreement with $M_V(0.9) = 7.12$. 

\item The globular cluster NGC 6752 has [Fe/H] $ = -1.6$ and $E(B-V) =
0.04$. The expected luminosity at $(B-V)_0 = 0.9$ is $M_V(0.9) = 7.46$. Renzini
et al (1996) report a $V$ band distance modulus of 13.05, based on the fit to
the white dwarf sequence on deep HST images.  Reading from the colour magnitude
diagram in Penny and Dickens (1986), the main sequence at $B-V = 1.94$ is at $V
\approx 20.6$, leading to an absolute magnitude of $M_V = 7.4$, which is in
good agreement with $M_V(0.9) = 7.46$.

\end{itemize} 

\begin{figure}
\epsfig{file=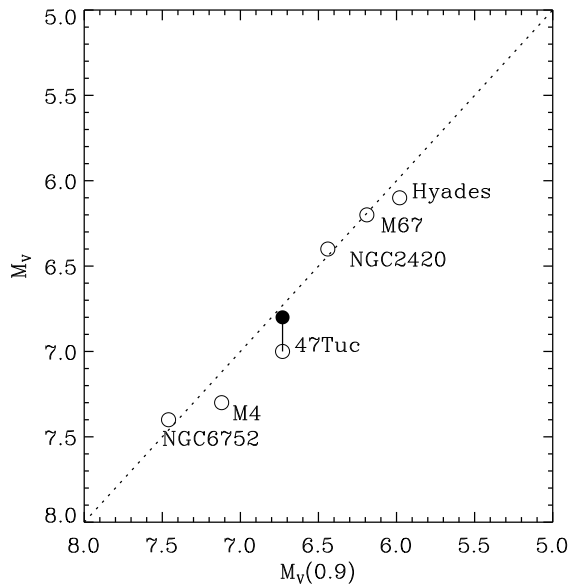,width=84mm}
\caption{Observed absolute magnitudes $M_V$ of main sequences in open and
globular clusters at $(B-V)_0 = 0.90$ versus the absolute magnitude $M_V(0.9)$
computed using Eq.~1. The data are shown in table 5. The line is the 1:1
relation. All clusters except 47 Tuc are a good fit. The open circle for 47 Tuc
is based on a distance modulus based on its white dwarf cooling sequence. The
filled circle for 47 Tuc is based on a more traditional distance modulus from the
position of the horizontal branch.}
\label{VversusV}
\end{figure}

The above calculations are summarised in Table 5.  Note that as far as possible
we have used distance moduli which are independent of main sequence fitting,
in order to avoid adopting moduli which are ultimately based on the same (or
similar) subdwarfs to those studied here. Overall the agreement between the
observed luminosity of the main sequences at $(B-V)_0 = 0.9$ and the value
obtained from Eq.~1 is satisfactory; a more detailed study would be
interesting.  One should note that at very low metallicities, the
metallicity-luminosity relation may not continue in the linear fashion shown in
panel (b) of Figure \ref{CMDkdwarfs}. More K dwarfs with high (super-solar) and
intermediate ($-1.2 < $[Fe/H] $< -0.5$) and Hipparcos parallaxes would be of
great interest in this regard, to check the linearity of the relation. There
are presently no K subdwarfs with good parallaxes and metallicities below
[Fe/H] $ \approx -2$ with which to check the relation at low metallicity.
Figure \ref{VversusV} shows the observed absolute magnitudes $M_V$ at $(B-V)_0
= 0.90$ versus the value $M_V(0.9)$ computed using Eq.~1.

\subsection{Usefulness of the relation}

The tightness of the metallicity-luminosity relation for K dwarfs is a
remarkable result for which there would be a wide range of applications. For
example:

\begin{itemize}

\item {\it Galactic chemical evolution:} For all K dwarfs in the Hipparcos
colour magnitude diagram, we can derive a metallicity with an accuracy of
better than 0.1 dex, and thus a very clean and complete determination of the
local metallicity distribution (for comparison with models of Galactic chemical
evolution). This work will appear in a companion paper (Kotoneva et al 2002).

\item {\it New distance indicator:} K dwarfs can be used as a distance
indicator, since we can determine quite accurately the absolute magnitude of K
dwarfs of a given colour and metallicity using Eq.~1, at least for
metallicities [Fe/H] $\ga -2.0$. K dwarfs could be used to determine a new set
of distance moduli to open clusters, the Galactic bulge, dwarf spheroidal
galaxies, the Magellanic clouds and most globular clusters, since for all of
these we may have an independent metallicity estimate.

\item {\it Galactic structure:} In a related study (Flynn, Holmberg and Lynch
2002, in preparation) we have found that using broadband $BVI$ photometry, we
can yield metallicity estimates for K dwarfs of similar accuracy (0.2 dex) to
the methods described in Kotoneva and Flynn (2002) (which use narrower band
Geneva or Str\"omgren filters). Thus, $BVI$ data could be used along various
lines of sight to (1) measure the structural parameters of the Galaxy, (2)
measure the local metallicity distribution at various sites in the Galaxy, such
as the inner and outer disk, the bulge region, along the Solar circle and
perpendicular to the Galactic plane. The metallicity distribution of stars as a
function of position in the Galactic disk would be an interesting constraint on
chemical evolution studies of the Galaxy.  Accurate distances would allow the
stellar density of K dwarfs (i.e. the luminosity/mass function) in the stellar
halo to be determined from simple star count data, and by extrapolating the
mass function of these K dwarfs, the density of remnants in the stellar halo
can be inferred.  Such remnants (in white dwarfs) might make up a fraction of
the Galactic dark matter (Oppenheimer et al 2001, Flynn et al 2002). We are
presently carrying out such a study (Flynn and Holmberg, 2002, in preparation).

A further obvious application is to the upcoming European Space Agency GAIA
mission which will measure $BVI$, parallax, proper motion and radial velocity data
for all K dwarfs within some 5-10 kpc.

\item {\it Helium:} Our calibration of the luminosity of K dwarfs as a function
of [Fe/H] can be used to test stellar models, which predict the luminosity more
generally as a function of helium abundance $Y$ and metals, $Z$. The use of
models in conjunction with this luminosity constraint might permit an improved
determination of the rate of change of helium with metals, $dY/dZ$ (Pagel and
Portinari, 1998).

\end{itemize}

\section{Conclusions}

We have shown that the luminosity of lower main sequence K dwarf stars at a
given colour is a simple function of metallicity ($\Delta M_V = 0.84375 \times
{\rm [Fe/H]} - 0.04577$). It was possible to show this because of the
availability of (1) accurate Hipparcos parallaxes, (2) a recently developed
photometric metallicity index for K dwarfs and (3) the possibility to identify
and remove the multiple stars from the sample. We have tested several sets of
isochrones from the literature and we show that none of them fit the data
completely, and in particular metal rich isochrones remain difficult to
construct.

We have adopted the solar isochrone from Jimenez, Flynn and Kotoneva (1998) as
a fiducial line relative to which the luminosity of K dwarfs as a function of a
metallicity has been empirically calibrated.  The calibration stars are a small
sample of K dwarfs with very accurate spectroscopically determined
metallicities and Hipparcos absolute magnitudes. We find that the relation is
very precise and may be used to derive metallicities for K dwarfs based on
their position in the colour-magnitude diagram with accuracies better than 0.1
dex. It is also a good distance indicator for K dwarfs if the metallicity can
be independently established.

\section*{Acknowledgments}
This research was supported by the Academy of Finland through its ANTARES
program for Space research and the Emil Aaltonen Foundation and Yrj{\"o}, Ville
and Kalle V\"ais\"al\"a Foundation.

\end{document}